%%%%%%%%%%%%%%%%%%  tex macros for preprints, cm version %%%%%%%%%%%%%%
%                     (P. Ginsparg, last updated 9/91)
%                if confused, type `b' in response to query 
%
%---------------------------------------------------------------------%
%% site dependent options: 
%% \unredoffs and \redoffs define horizontal and vertical offsets 
%% respectively for unreduced and reduced modes. \speclscape defines
%% the \special{} call that sets printer to landscape (sideways) mode.
%% from standard set below, leave uncommented as appropriate or redefine
%
%%% next 400dpi
%\def\unredoffs{} \def\redoffs{\voffset=-.31truein\hoffset=-.48truein}
%\def\speclscape{\special{landscape}}
%
%%% apple lw
\def\unredoffs{} 

%
%%% qms lasergrafix:
%\def\unredoffs{} \def\redoffs{\voffset=-.4truein\hoffset=.125truein}
%\def\speclscape{\special{qms: landscape}}
%
%%% saclay A4 paper:
%\def\unredoffs{\hoffset-.14truein\voffset-.2truein} 
%\def\redoffs{\voffset=-.45truein\hoffset=-.21truein} 
%\def\speclscape{\special{landscape}}
%
%---------------------------------------------------------------------%
%
\newbox\leftpage \newdimen\fullhsize \newdimen\hstitle \newdimen\hsbody
\tolerance=1000\hfuzz=2pt
\catcode`\@=11 % This allows us to modify PLAIN macros.
%HIER gefummelt: die bloede Entscheidungsfrage auskommentiert. 
%\def\bigans{b }
%\message{ big or little (b/l)? }   \read-1 to\answ    
%\ifx\answ\bigans\message{(This will come out unreduced.}
\magnification=1200\unredoffs\baselineskip=16pt plus 2pt minus 1pt
\hsbody=\hsize \hstitle=\hsize %take default values for unreduced format
%
%\else\message{(This will be reduced.} \let\l@r=L
%\magnification=1000\baselineskip=16pt plus 2pt minus 1pt \vsize=7truein
%\redoffs \hstitle=8truein\hsbody=4.75truein\fullhsize=10truein\hsize=\hsbody
%
%\output={\ifnum\pageno=0 %%% This is the HUTP version
%  \shipout\vbox{\speclscape{\hsize\fullhsize\makeheadline}
%    \hbox to \fullhsize{\hfill\pagebody\hfill}}\advancepageno
%  \else
%  \almostshipout{\leftline{\vbox{\pagebody\makefootline}}}\advancepageno 
%  \fi}
%\def\almostshipout#1{\if L\l@r \count1=1 \message{[\the\count0.\the\count1]}
%      \global\setbox\leftpage=#1 \global\let\l@r=R
% \else \count1=2
%  \shipout\vbox{\speclscape{\hsize\fullhsize\makeheadline}
%      \hbox to\fullhsize{\box\leftpage\hfil#1}}  \global\let\l@r=L\fi}
%\fi
%
%bis HIER naemlich
%
%---------------------------------------------------------------------
%
\newcount\yearltd\yearltd=\year\advance\yearltd by -1900

\def\Title#1#2{\nopagenumbers\abstractfont\hsize=\hstitle\rightline{#1}%
\vskip 1in\centerline{\titlefont #2}\abstractfont\vskip .5in\pageno=0}
\def\Date#1{\vfill\leftline{#1}\tenpoint\supereject\global\hsize=\hsbody%
\footline={\hss\tenrm\folio\hss}}%      restores pagenumbers
%
%       use following instead of \Date on the preliminary draft, 
%       puts date/time on each page in big mode, writes labels in margins

\def\draftmode{\message{ DRAFTMODE }\def\draftdate{{\rm preliminary draft:
\number\month/\number\day/\number\yearltd\ \ \hourmin}}%
\headline={\hfil\draftdate}\writelabels\baselineskip=20pt plus 2pt minus 2pt
 {\count255=\time\divide\count255 by 60 \xdef\hourmin{\number\count255}
  \multiply\count255 by-60\advance\count255 by\time
  \xdef\hourmin{\hourmin:\ifnum\count255<10 0\fi\the\count255}}}
%       use \nolabels to get rid of eqn, ref, and fig labels in draft mode
\def\nolabels{\def\wrlabeL##1{}\def\eqlabeL##1{}\def\reflabeL##1{}}
\def\writelabels{\def\wrlabeL##1{\leavevmode\vadjust{\rlap{\smash%
{\line{{\escapechar=` \hfill\rlap{\sevenrm\hskip.03in\string##1}}}}}}}%
\def\eqlabeL##1{{\escapechar-1\rlap{\sevenrm\hskip.05in\string##1}}}%
\def\reflabeL##1{\noexpand\llap{\noexpand\sevenrm\string\string\string##1}}}
\nolabels
%
% tagged sec numbers
\global\newcount\secno \global\secno=0
\global\newcount\meqno \global\meqno=1
\def\newsec#1{\global\advance\secno by1\message{(\the\secno. #1)}
%\ifx\answ\bigans \vfill\eject \else \bigbreak\bigskip \fi  %if desired
\global\subsecno=0\eqnres@t\noindent{\bf\the\secno. #1}
\writetoca{{\secsym} {#1}}\par\nobreak\medskip\nobreak}
\def\eqnres@t{\xdef\secsym{\the\secno.}\global\meqno=1\bigbreak\bigskip}
\def\sequentialequations{\def\eqnres@t{\bigbreak}}\xdef\secsym{}
\global\newcount\subsecno \global\subsecno=0
\def\subsec#1{\global\advance\subsecno by1\message{(\secsym\the\subsecno. #1)}
\ifnum\lastpenalty>9000\else\bigbreak\fi
\noindent{\it\secsym\the\subsecno. #1}\writetoca{\string\quad 
{\secsym\the\subsecno.} {#1}}\par\nobreak\medskip\nobreak}
\def\appendix#1#2{\global\meqno=1\global\subsecno=0\xdef\secsym{\hbox{#1.}}
\bigbreak\bigskip\noindent{\bf Appendix #1. #2}\message{(#1. #2)}
\writetoca{Appendix {#1.} {#2}}\par\nobreak\medskip\nobreak}
%
%       \eqn\label{a+b=c}       gives displayed equation, numbered
%                               consecutively within sections.
%     \eqnn and \eqna define labels in advance (of eqalign?)
%
\def\eqnn#1{\xdef #1{(\secsym\the\meqno)}\writedef{#1\leftbracket#1}%
\global\advance\meqno by1\wrlabeL#1}
\def\eqna#1{\xdef #1##1{\hbox{$(\secsym\the\meqno##1)$}}
\writedef{#1\numbersign1\leftbracket#1{\numbersign1}}%
\global\advance\meqno by1\wrlabeL{#1$\{\}$}}
\def\eqn#1#2{\xdef #1{(\secsym\the\meqno)}\writedef{#1\leftbracket#1}%
\global\advance\meqno by1$$#2\eqno#1\eqlabeL#1$$}
%
%                        footnotes
\newskip\footskip\footskip14pt plus 1pt minus 1pt %sets footnote baselineskip
\def\footnotefont{\ninepoint}\def\f@t#1{\footnotefont #1\@foot}
\def\f@@t{\baselineskip\footskip\bgroup\footnotefont\aftergroup\@foot\let\next}
\setbox\strutbox=\hbox{\vrule height9.5pt depth4.5pt width0pt}
\global\newcount\ftno \global\ftno=0
\def\foot{\global\advance\ftno by1\footnote{$^{\the\ftno}$}}
%
%say \footend to put footnotes at end
%will cause problems if \ref used inside \foot, instead use \nref before
\newwrite\ftfile   
\def\footend{\def\foot{\global\advance\ftno by1\chardef\wfile=\ftfile
$^{\the\ftno}$\ifnum\ftno=1\immediate\openout\ftfile=foots.tmp\fi%
\immediate\write\ftfile{\noexpand\smallskip%
\noexpand\item{f\the\ftno:\ }\pctsign}\findarg}%
\def\footatend{\vfill\eject\immediate\closeout\ftfile{\parindent=20pt
\centerline{\bf Footnotes}\nobreak\bigskip\input foots.tmp }}}
\def\footatend{}
%
%     \ref\label{text}
% generates a number, assigns it to \label, generates an entry.
% To list the refs on a separate page,  \listrefs
%
\global\newcount\refno \global\refno=1
\newwrite\rfile
\def\ref{[\the\refno]\nref}
\def\nref#1{\xdef#1{[\the\refno]}\writedef{#1\leftbracket#1}%
\ifnum\refno=1\immediate\openout\rfile=refs.tmp\fi
\global\advance\refno by1\chardef\wfile=\rfile\immediate
\write\rfile{\noexpand\item{#1\ }\reflabeL{#1\hskip.31in}\pctsign}\findarg}
%       horrible hack to sidestep tex \write limitation
\def\findarg#1#{\begingroup\obeylines\newlinechar=`\^^M\pass@rg}
{\obeylines\gdef\pass@rg#1{\writ@line\relax #1^^M\hbox{}^^M}%
\gdef\writ@line#1^^M{\expandafter\toks0\expandafter{\striprel@x #1}%
\edef\next{\the\toks0}\ifx\next\em@rk\let\next=\endgroup\else\ifx\next\empty%
\else\immediate\write\wfile{\the\toks0}\fi\let\next=\writ@line\fi\next\relax}}
\def\striprel@x#1{} \def\em@rk{\hbox{}} 
\def\lref{\begingroup\obeylines\lr@f}
\def\lr@f#1#2{\gdef#1{\ref#1{#2}}\endgroup\unskip}

\def\addref#1{\immediate\write\rfile{\noexpand\item{}#1}} %now unnecessary
\def\startrefs#1{\immediate\openout\rfile=refs.tmp\refno=#1}
\def\xref{\expandafter\xr@f}\def\xr@f[#1]{#1}
\def\refs#1{\count255=1[\r@fs #1{\hbox{}}]}
\def\r@fs#1{\ifx\und@fined#1\message{reflabel \string#1 is undefined.}%
\nref#1{need to supply reference \string#1.}\fi%
\vphantom{\hphantom{#1}}\edef\next{#1}\ifx\next\em@rk\def\next{}%
\else\ifx\next#1\ifodd\count255\relax\xref#1\count255=0\fi%
\else#1\count255=1\fi\let\next=\r@fs\fi\next}
%

%
% this is ugly, but moore insists
\newwrite\ffile\global\newcount\figno \global\figno=1
\def\fig{fig.~\the\figno\nfig}
\def\nfig#1{\xdef#1{fig.~\the\figno}%
\writedef{#1\leftbracket fig.\noexpand~\the\figno}%
\ifnum\figno=1\immediate\openout\ffile=figs.tmp\fi\chardef\wfile=\ffile%
\immediate\write\ffile{\noexpand\medskip\noexpand\item{Fig.\ \the\figno. }
\reflabeL{#1\hskip.55in}\pctsign}\global\advance\figno by1\findarg}
\def\xfig{\expandafter\xf@g}\def\xf@g fig.\penalty\@M\ {}
\def\figs#1{figs.~\f@gs #1{\hbox{}}}
\def\f@gs#1{\edef\next{#1}\ifx\next\em@rk\def\next{}\else
\ifx\next#1\xfig #1\else#1\fi\let\next=\f@gs\fi\next}
\newwrite\lfile
{\escapechar-1\xdef\pctsign{\string\%}\xdef\leftbracket{\string\{}
\xdef\rightbracket{\string\}}\xdef\numbersign{\string\#}}

\def\writestop{\def\writestoppt{\immediate\write\lfile{\string\pageno%
\the\pageno\string\startrefs\leftbracket\the\refno\rightbracket%
\string\def\string\secsym\leftbracket\secsym\rightbracket%
\string\secno\the\secno\string\meqno\the\meqno}\immediate\closeout\lfile}}
\def\writestoppt{}\def\writedef#1{}
\def\seclab#1{\xdef #1{\the\secno}\writedef{#1\leftbracket#1}\wrlabeL{#1=#1}}
\def\subseclab#1{\xdef #1{\secsym\the\subsecno}%
\writedef{#1\leftbracket#1}\wrlabeL{#1=#1}}
\newwrite\tfile \def\writetoca#1{}
\def\leaderfill{\leaders\hbox to 1em{\hss.\hss}\hfill}
%       use this to write file with table of contents
\def\writetoc{\immediate\openout\tfile=toc.tmp 
   \def\writetoca##1{{\edef\next{\write\tfile{\noindent ##1 
   \string\leaderfill {\noexpand\number\pageno} \par}}\next}}}
%       and this lists table of contents on second pass

%
\catcode`\@=12 % at signs are no longer letters
%
%       Unpleasantness in calling in abstract and title fonts
\edef\tfontsize{\ifx\answ\bigans scaled\magstep3\else scaled\magstep4\fi}
\font\titlerm=cmr10 \tfontsize \font\titlerms=cmr7 \tfontsize
\font\titlermss=cmr5 \tfontsize \font\titlei=cmmi10 \tfontsize
\font\titleis=cmmi7 \tfontsize \font\titleiss=cmmi5 \tfontsize
\font\titlesy=cmsy10 \tfontsize \font\titlesys=cmsy7 \tfontsize
\font\titlesyss=cmsy5 \tfontsize \font\titleit=cmti10 \tfontsize
\skewchar\titlei='177 \skewchar\titleis='177 \skewchar\titleiss='177
\skewchar\titlesy='60 \skewchar\titlesys='60 \skewchar\titlesyss='60
\def\titlefont{\def\rm{\fam0\titlerm}% switch to title font
\textfont0=\titlerm \scriptfont0=\titlerms \scriptscriptfont0=\titlermss
\textfont1=\titlei \scriptfont1=\titleis \scriptscriptfont1=\titleiss
\textfont2=\titlesy \scriptfont2=\titlesys \scriptscriptfont2=\titlesyss
\textfont\itfam=\titleit \def\it{\fam\itfam\titleit}\rm}
 \ifx\answ\bigans\else scaled\magstep1\fi
\ifx\answ\bigans\def\abstractfont{\tenpoint}\else
\font\abssl=cmsl10 scaled \magstep1
\font\absrm=cmr10 scaled\magstep1 \font\absrms=cmr7 scaled\magstep1
\font\absrmss=cmr5 scaled\magstep1 \font\absi=cmmi10 scaled\magstep1
\font\absis=cmmi7 scaled\magstep1 \font\absiss=cmmi5 scaled\magstep1
\font\abssy=cmsy10 scaled\magstep1 \font\abssys=cmsy7 scaled\magstep1
\font\abssyss=cmsy5 scaled\magstep1 \font\absbf=cmbx10 scaled\magstep1
\skewchar\absi='177 \skewchar\absis='177 \skewchar\absiss='177
\skewchar\abssy='60 \skewchar\abssys='60 \skewchar\abssyss='60
\def\abstractfont{\def\rm{\fam0\absrm}% switch to abstract font
\textfont0=\absrm \scriptfont0=\absrms \scriptscriptfont0=\absrmss
\textfont1=\absi \scriptfont1=\absis \scriptscriptfont1=\absiss
\textfont2=\abssy \scriptfont2=\abssys \scriptscriptfont2=\abssyss
\textfont\itfam=\bigit \def\it{\fam\itfam\bigit}\def\footnotefont{\tenpoint}%
\textfont\slfam=\abssl \def\sl{\fam\slfam\abssl}%
\textfont\bffam=\absbf \def\bf{\fam\bffam\absbf}\rm}\fi
\def\tenpoint{\def\rm{\fam0\tenrm}% switch back to 10-point type
\textfont0=\tenrm \scriptfont0=\sevenrm \scriptscriptfont0=\fiverm
\textfont1=\teni  \scriptfont1=\seveni  \scriptscriptfont1=\fivei
\textfont2=\tensy \scriptfont2=\sevensy \scriptscriptfont2=\fivesy
\textfont\itfam=\tenit \def\it{\fam\itfam\tenit}\def\footnotefont{\ninepoint}%
\textfont\bffam=\tenbf \def\bf{\fam\bffam\tenbf}\def\sl{\fam\slfam\tensl}\rm}
\font\ninerm=cmr9 \font\sixrm=cmr6 \font\ninei=cmmi9 \font\sixi=cmmi6 
\font\ninesy=cmsy9 \font\sixsy=cmsy6 \font\ninebf=cmbx9 
\font\nineit=cmti9 \font\ninesl=cmsl9 \skewchar\ninei='177
\skewchar\sixi='177 \skewchar\ninesy='60 \skewchar\sixsy='60 
\def\ninepoint{\def\rm{\fam0\ninerm}% switch to footnote font
\textfont0=\ninerm \scriptfont0=\sixrm \scriptscriptfont0=\fiverm
\textfont1=\ninei \scriptfont1=\sixi \scriptscriptfont1=\fivei
\textfont2=\ninesy \scriptfont2=\sixsy \scriptscriptfont2=\fivesy
\textfont\itfam=\ninei \def\it{\fam\itfam\nineit}\def\sl{\fam\slfam\ninesl}%
\textfont\bffam=\ninebf \def\bf{\fam\bffam\ninebf}\rm} 
%
%---------------------------------------------------------------------
%
\def\noblackbox{\overfullrule=0pt}
\hyphenation{anom-aly anom-alies coun-ter-term coun-ter-terms}
\def\inv{^{\raise.15ex\hbox{${\scriptscriptstyle -}$}\kern-.05em 1}}

\def\Dsl{\,\raise.15ex\hbox{/}\mkern-13.5mu D} %this one can be subscripted
\def\dsl{\raise.15ex\hbox{/}\kern-.57em\partial}

\font\bigit=cmti10 scaled \magstep1
 %pound sterling
\def\lspace{\ifx\answ\bigans{}\else\qquad\fi}
\def\lbspace{\ifx\answ\bigans{}\else\hskip-.2in\fi} % $$\lbspace...$$
\def\boxeqn#1{\vcenter{\vbox{\hrule\hbox{\vrule\kern3pt\vbox{\kern3pt
        \hbox{${\displaystyle #1}$}\kern3pt}\kern3pt\vrule}\hrule}}}
\def\mbox#1#2{\vcenter{\hrule \hbox{\vrule height#2in
                \kern#1in \vrule} \hrule}}  %e.g. \mbox{.1}{.1}
%       matters of taste
%\def\tilde{\widetilde} \def\bar{\overline} \def\hat{\widehat}
%
% some sample definitions
  %     curly letters

\def\vev#1{\langle #1 \rangle}

\def\darr#1{\raise1.5ex\hbox{$\leftrightarrow$}\mkern-16.5mu #1}
 %pound sterling

\def\half{{\textstyle{1\over2}}} %puts a small half in a displayed eqn
\def\roughly#1{\raise.3ex\hbox{$#1$\kern-.75em\lower1ex\hbox{$\sim$}}}

\openup -1pt
\input epsf
\expandafter\ifx\csname pre amssym.tex at\endcsname\relax \else\endinput\fi
\expandafter\chardef\csname pre amssym.tex at\endcsname=\the\catcode`\@
\catcode`\@=11
\ifx\undefined\newsymbol \else \begingroup\def\input#1 {\endgroup}\fi
\expandafter\ifx\csname amssym.def\endcsname\relax \else\endinput\fi
\expandafter\edef\csname amssym.def\endcsname{%
       \catcode`\noexpand\@=\the\catcode`\@\space}
\catcode`\@=11
\def\undefine#1{\let#1\undefined}
\def\newsymbol#1#2#3#4#5{\let\next@\relax
 \ifnum#2=\@ne\let\next@\msafam@\else
 \ifnum#2=\tw@\let\next@\msbfam@\fi\fi
 \mathchardef#1="#3\next@#4#5}
\def\mathhexbox@#1#2#3{\relax
 \ifmmode\mathpalette{}{\m@th\mathchar"#1#2#3}%
 \else\leavevmode\hbox{$\m@th\mathchar"#1#2#3$}\fi}
\def\hexnumber@#1{\ifcase#1 0\or 1\or 2\or 3\or 4\or 5\or 6\or 7\or 8\or
 9\or A\or B\or C\or D\or E\or F\fi}
\font\tenmsa=msam10
\font\sevenmsa=msam7
\font\fivemsa=msam5
\newfam\msafam
\textfont\msafam=\tenmsa
\scriptfont\msafam=\sevenmsa
\scriptscriptfont\msafam=\fivemsa
\edef\msafam@{\hexnumber@\msafam}
\mathchardef\dabar@"0\msafam@39
\def\maltese{{\mathhexbox@\msafam@7A}}
\font\tenmsb=msbm10
\font\sevenmsb=msbm7
\font\fivemsb=msbm5
\newfam\msbfam
\textfont\msbfam=\tenmsb
\scriptfont\msbfam=\sevenmsb
\scriptscriptfont\msbfam=\fivemsb
\edef\msbfam@{\hexnumber@\msbfam}
\def\Bbb#1{{\fam\msbfam\relax#1}}
\def\widehat#1{\setbox\z@\hbox{$\m@th#1$}%
 \ifdim\wd\z@>\tw@ em\mathaccent"0\msbfam@5B{#1}%
 \else\mathaccent"0362{#1}\fi}
\def\widetilde#1{\setbox\z@\hbox{$\m@th#1$}%
 \ifdim\wd\z@>\tw@ em\mathaccent"0\msbfam@5D{#1}%
 \else\mathaccent"0365{#1}\fi}
\font\teneufm=eufm10
\font\seveneufm=eufm7
\font\fiveeufm=eufm5
\newfam\eufmfam
\textfont\eufmfam=\teneufm
\scriptfont\eufmfam=\seveneufm
\scriptscriptfont\eufmfam=\fiveeufm

\csname amssym.def\endcsname
\relax
\newsymbol\smallsetminus 2272
\noblackbox
%%% Paragraphs
\newcount\figno
\figno=0
\def\mathrm#1{{\rm #1}}
\def\fig#1#2#3{
\par\begingroup\parindent=0pt\leftskip=1cm\rightskip=1cm\parindent=0pt
\baselineskip=11pt
\global\advance\figno by 1
\midinsert
\epsfxsize=#3
\centerline{\epsfbox{#2}}
\vskip 12pt
\centerline{{\bf Figure \the\figno} #1}\par
\endinsert\endgroup\par}
\font\tenmsb=msbm10       \font\sevenmsb=msbm7
\font\fivemsb=msbm5       \newfam\msbfam
\textfont\msbfam=\tenmsb  \scriptfont\msbfam=\sevenmsb
\scriptscriptfont\msbfam=\fivemsb
\def\Bbb#1{{\fam\msbfam\relax#1}}

\def\Rop{{\Bbb R}}
\def\Zop{{\Bbb Z}}
\def\Cop{{\Bbb C}}

\def\bbbc{{\mathchoice {\setbox0=\hbox{$\displaystyle\rm C$}\hbox{\hbox
to0pt{\kern0.4\wd0\vrule height0.9\ht0\hss}\box0}}
{\setbox0=\hbox{$\textstyle\rm C$}\hbox{\hbox
to0pt{\kern0.4\wd0\vrule height0.9\ht0\hss}\box0}}
{\setbox0=\hbox{$\scriptstyle\rm C$}\hbox{\hbox
to0pt{\kern0.4\wd0\vrule height0.9\ht0\hss}\box0}}
{\setbox0=\hbox{$\scriptscriptstyle\rm C$}\hbox{\hbox
to0pt{\kern0.4\wd0\vrule height0.9\ht0\hss}\box0}}}}
\def\figlabel#1{\xdef#1{\the\figno}}
\def\pano{\par\noindent}

%%% special math symbols
\def\pmb#1{\setbox0=\hbox{#1}%
 \kern-.025em\copy0\kern-\wd0
 \kern.05em\copy0\kern-\wd0
 \kern-.025em\raise.0433em\box0 }

\def\half{{1\over 2}}

%\def\Cop{\relax\,\hbox{$\kern-.3em{\rm C}$}}
%\def\Cop{\relax\,\hbox{$\inbar\kern-.3em{\rm C}$}}
%\def\Rop{\relax{\rm I\kern-.18em R}}
%\def\Nop{\relax{\rm I\kern-.18em N}}
%\def\Pop{\relax{\rm I\kern-.18em P}}
%\def\Zop{\rlx\leavevmode\ifmmode\mathchoice{\hbox{\cmss Z\kern-.4em Z}}
% {\hbox{\cmss Z\kern-.4em Z}}{\lower.9pt\hbox{\cmsss Z\kern-.36em Z}}
% {\lower1.2pt\hbox{\cmsss Z\kern-.36em Z}}\else{\cmss Z\kern-.4em
% Z}\fi}

%%% misc.

\def\H{{\cal H}}

\def\A{{\cal{A}}}

\def\B#1,#2,#3.{{}^{(#1)}\!B_{#2}^{#3}}

\def\One{{\rm 1\!\!1}}

\def\qt{{\tilde q}}

\def\ss{\scriptstyle}
\def\cb#1,#2,#3,#4,#5.{{f^{#1}\!%
\left[\ss{{\ss#3\,#4}\atop{\ss#2\,#5}}\right]}}
\def\fm#1,#2,#3,#4,#5,#6.{{F_{#1#2}\!%
\left[\ss{{\ss#4\,#5}\atop{\ss#3\,#6}}\right]}}

\def\vec#1{{\vert\, #1\, \rangle}}
\def\ivec#1{{\vert\, #1\, \rangle\!\rangle}}
\def\iivec#1{{\vert\!\vert\, #1\, \rangle\!\rangle}}

\def\cev#1{{\langle \,#1\, \vert}}
\def\icev#1{{\langle\!\langle \,#1\, \vert}}
\def\iicev#1{{\langle\!\langle \,#1\, \vert\!\vert}}

\def\cevec#1#2{{\langle \,#1\, \vert \,#2\, \rangle}}

\def\vev#1{{\langle\,#1\,\rangle}}

\def\gvec{\iivec{ g }}

\def\blank#1{}
\def\ie{{\it i.e.}}

\def\hsmallsetminus{\hbox{\raise1.5pt\hbox{$\smallsetminus$}}}
\def\su{\hbox{$\widehat{\rm su}$}}
\def\Vir{\hbox{\sl Vir}}
%%% further macros

%%% References

\lref\cklm{C.G.\ Callan, I.R.\ Klebanov, A.W.\ Ludwig and
J.M.\ Maldacena, {\it Exact solution of a boundary conformal field
theory}, Nucl.\ Phys.\ {\bf B422}, 417 (1994); {\tt hep-th/9402113}.} 

\lref\cardy{J.L.\ Cardy, {\it Boundary conditions, fusion rules and
the Verlinde formula}, Nucl.\ Phys.\ {\bf B324}, 581 (1989).}

\lref\fk{I.B.\ Frenkel and V.G.\ Kac, {\it Basic representations of
affine Lie algebras and dual resonance models}, Invent.\ Math.\ 
{\bf 62}, 23 (1981).} 

\lref\fs{J.\ Fuchs and C.\ Schweigert, {\it A classifying algebra for
boundary conditions}, Phys.\ Lett.\ {\bf B414}, 251 (1997);
{\tt hep-th/9708141}. {\it Category theory for conformal boundary
conditions}, {\tt math.CT/0106050}.}

\lref\grw{K.\ Graham, I.\ Runkel and G.M.T.\ Watts,
{\it Minimal model boundary flows and $c=1$ CFT}, 
{\tt hep-th/0101187}, Nucl.\ Phys.\ {\bf B}, {\it in press}.} 

\lref\ham{M.\ Hamermesh, {\it Group theory and its applications to 
physical problems}, Addison-Wesley (1962).}

\lref\ishi{N.\ Ishibashi, {\it The boundary and crosscap states in
conformal field theories}, Mod.\ Phys.\ Lett.\ {\bf A4}, 251 (1989).}

\lref\klebpol{I.R.\ Klebanov and A.M.\ Polyakov, {\it Interaction of 
discrete states in two-dimensional string theory}, Mod.\ Phys.\ Lett.\
{\bf A6}, 3273 (1991); {\tt hep-th/9109032}.} 

\lref\lew{D.C.\ Lewellen, {\it Sewing constraints for conformal field
theories on surfaces with boundaries}, Nucl.\ Phys.\ {\bf B372}, 654
(1992).} 

\lref\polthor{J.\ Polchinski and L.\ Thorlacius, {\it Free fermion
representation of a boundary conformal field theory}, Phys.\ Rev.\ 
{\bf D50}, 622 (1994); {\tt hep-th/9404008}.}

\lref\rsone{A.\ Recknagel and V.\ Schomerus, {\it Boundary deformation  
theory and moduli spaces of D-branes}, Nucl.\ Phys.\ {\bf B545}, 233 
(1999); {\tt hep-th/9811237}.}

\lref\segal{G.B.\ Segal, {\it Unitary representations of some infinite
dimensional groups}, Commun.\ Math.\ Phys.\ {\bf 80}, 301 (1981).} 

\lref\knapp{A.W.\ Knapp, {\it Representation theory of semisimple
groups: an overview based on examples}, Princeton University Press,
Princeton (1986).}
 
\lref\GG{M.B.\ Green and M.\ Gutperle, {\it Symmetry breaking at 
enhanced symmetry points}, Nucl.\ Phys.\ {\bf B460}, 77 (1996); 
{\tt hep-th/9509171}.}

\lref\friedan{D. Friedan, {\it The space of conformal boundary
conditions for the $c=1$ Gaussian model}, unpublished note (1999).} 

\lref\pss{G.\ Pradisi, A.\ Sagnotti and Y.S.\ Stanev, 
{\it Completeness  conditions for boundary operators in 2d conformal
field theory}, Phys.\ Lett.\ {\bf B381}, 97 (1996); 
{\tt hep-th/9603097}.} 

\lref\bppz{R.E.\ Behrend, P.A.\ Pearce, V.B.\ Petkova and J.-B.\ Zuber,  
{\it Boundary conditions in rational conformal field theories}, 
Nucl.\ Phys.\ {\bf B579}, 707 (2000); {\tt hep-th/9908036}.}

\lref\pz{V.B.\ Petkova and J.-B.\ Zuber, {\it The many faces of Ocneanu
cells}, Nucl.\ Phys.\ {\bf B603}, 449 (2001); {\tt hep-th/0101151}.}

\lref\ms{G.\ Moore and N.\ Seiberg, {\it Classical and quantum
conformal field theory}, Commun.\ Math.\ Phys.\ {\bf 123}, 177
(1989).} 

\lref\mms{J.\ Maldacena, G.\ Moore and N.\ Seiberg, {\it  Geometrical 
interpretation of D-branes in gauged WZW models}, 
{\tt hep-th/0105038}.} 

\lref\friedan{D.\ Friedan, {\it The space of conformal boundary
conditions for the $c=1$ Gaussian model}, unpublished note (1999).} 

\lref\fffs{G.\ Felder, J.\ Fr\"ohlich, J.\ Fuchs and C.\ Schweigert,
{\it Correlation functions and boundary conditions in RCFT and
three-dimensional topology}, {\tt hep-th/9912239}.}

\lref\polcai{J.\ Polchinski, Y.\ Cai, {\it Consistency of open 
superstring theories}, Nucl.\ Phys.\ {\bf B296}, 91 (1988).}

\lref\rsgep{A.\ Recknagel, V.\ Schomerus, {\it D-branes in 
Gepner models}, Nucl.\ Phys.\ {\bf B531}, 185 (1998);
{\tt hep-th/9712186}.}

%%% Title page
\Title{\vbox{
\hbox{hep--th/0108102}
\hbox{KCL-MTH-01-36}}}
{\vbox{\centerline{
The conformal boundary states for SU(2) at level 1}
}}
\centerline{M.R.\ Gaberdiel\footnote{}{\hskip-14pt{\tt e-mails:
mrg@mth.kcl.ac.uk, anderl@mth.kcl.ac.uk, gmtw@mth.kcl.ac.uk}},
$\;\,$A.\ Recknagel$\;\,$ and$\;\,$ G.M.T.\ Watts
}
\bigskip
\centerline{\it Department of Mathematics, King's College London}
\centerline{\it Strand, London WC2R 2LS, U.K.}
\smallskip
\vskip2cm
\centerline{\bf Abstract}
\bigskip

{\narrower
\noindent For the case of the SU(2) WZW model at level one, the boundary 
states that only preserve the conformal symmetry are analysed. Under the
assumption that the usual Cardy boundary states as well as their
marginal deformations are consistent, the most general conformal
boundary states are determined. They are found to be parametrised by
group elements in SL$(2,\Cop)$.

}

\bigskip
\Date{\hbox{\tt \ \ August 2001}}
%%% text

\newsec{Introduction}

Much has been understood in the last few years about conformal field
theories on surfaces with a boundary. In particular, if the conformal 
field theory is rational with respect to some chiral algebra, the
boundary states that preserve this symmetry algebra have been
classified at least for a large class of cases
\refs{\cardy,\bppz}. However, much less is known about boundary 
conditions that only preserve a symmetry algebra with respect to which
the original theory is not rational.

In many applications, including string theory, one is ultimately
interested in boundary conditions that preserve the conformal symmetry
but not necessarily any larger symmetry.  The only conformal field
theories which are rational with respect to the Virasoro algebra alone
are the minimal models. All other conformal field theories are at best
rational with respect to some larger symmetry algebra ${\cal W}$ (but
not with respect to $\Vir \subset {\cal W}$). The physically important
boundary conditions may then break the symmetry to \Vir. In this paper
we construct such boundary states for a simple example, the SU(2) WZW
model at level $1$. This theory is rational with respect to the affine
$\su(2)$ symmetry, but not with respect to the Virasoro algebra. We
will arrive at results that are similar, but not identical, to claims
made by Friedan in unpublished work \refs{\friedan}; see Section~5 for
further comments.  

In order to obtain constraints on the possible conformal boundary
states of this model we make use of one of the sewing relations of
\refs{\lew}, the so-called factorisation (or cluster) condition. As
we explain in detail in Section~2, this constraint requires that
certain bulk-boundary structure constants satisfy a set of non-linear
equations (sometimes referred to as the classifying algebra in this
context). The coefficients in these equations are in principle 
determined in terms of bulk operator product coefficients and fusing
matrices. For rational theories these are sometimes available, but for
the case at hand (where we take the chiral algebra to be the Virasoro
algebra alone and therefore deal with a non-rational theory) they are
not known explicitly. We therefore use an indirect argument to
determine the coefficients appearing in the cluster condition. To this
end, we observe that these coefficients can be deduced from the
knowledge of a sufficiently large class of solutions. In our case, the
latter arise from the boundary states that preserve the full affine
symmetry; these have been constructed before (see for example
\refs{\GG,\rsone}), and are believed to be consistent. In fact, they
can be obtained from the Cardy solution \refs{\cardy} by marginal
deformations. 

Having determined the cluster condition explicitly we then classify
all its solutions (\ie\ all the one-dimensional representations of
the classifying algebra), and therefore all possible 
(fundamental) D-branes of the theory. As it turns out, the most
general D-branes are parametrised by group elements in SL$(2,\Cop)$.
On the level of our analysis we cannot prove that all of these
D-branes are actually consistent (\ie\ satisfy all remaining sewing
constraints as well), but our arguments imply that the theory does not
have any other (fundamental) D-branes. This is quite a surprising
result since one may have thought that the space of D-branes that only
preserve the conformal symmetry would be much larger.

In Section~4 we then check that this larger class of boundary
conditions satisfies Cardy's condition. In particular, we find that
the spectrum of open strings between two such D-branes organises
itself into representations of the Virasoro algebra; however, in
general, the conformal weights that appear are complex (rather than 
real). We conclude in Section~5 with some comments about the
interpretation of these D-branes.

\newsec{Constraints on  boundary conditions for 
$\widehat{\bf su}{\bf (2)}_{k=1}$}  

Let us begin by recalling some properties of the theory on the plane
(the bulk theory) which in our case is the SU(2) WZW model at level
$k=1$. This theory has left- and right-moving currents $J^a(z)$ and
$\bar J^a(\bar z)$ whose modes $J^a_n$ and $\bar J^a_n$ define two
commuting copies of the affine algebra $\widehat{\rm su}(2)_1$. The
space of states of the bulk theory $\H_{\rm bulk}$ can therefore be
decomposed into highest weight representations of these two algebras. 

There are two highest weight representations of the affine algebra
$\su(2)$ at level $1$ that actually define representations of the
conformal field theory (or the corresponding vertex operator algebra):
the vacuum representation $\H_0$ that is generated from a state
transforming in the singlet representations of the horizontal su(2)
algebra (spanned by the zero modes $J^a_0$), and the representation
$\H_{\half}$ for which the highest weight states transform in the
doublet representation of the horizontal su(2) algebra. For $k=1$
there is only one modular invariant combination of these
representations and thus a unique bulk theory, whose space of states
is of the form
\eqn\bulkh{ \H_{\rm bulk} = \left( \H^{\widehat{\rm su}(2)}_0
\otimes {\bar\H^{\widehat{\rm su}(2)}_0} \right) \; \oplus \; \left(
\H^{\widehat{\rm su}(2)}_{\half} \otimes
{\bar\H^{\widehat{\rm su}(2)}_{\half}}\right) \,.}  
Because of the Sugawara construction, every highest weight
representation of $\su(2)_1$ also forms a representation of the
Virasoro algebra \Vir\ with $c=1$. The generators of the Virasoro
algebra commute with the current zero modes, and thus the
$\su(2)$ representations can be decomposed into representations
of su(2)$\;\oplus\;\Vir$. If we denote the $(2j+1)$-dimensional spin
$j$ representation of su(2) by $V^j$ and the Virasoro algebra
irreducible highest weight representation of weight $h$ by 
$\H^{\rm Vir}_h$, then we have 
\eqn\decompa{
  \H^{\widehat{\rm su}(2)}_j
   = \sum_{n =0}^\infty  V^{(n+j)} \otimes \H^{\rm Vir}_{(n+j)^2} \;.
}
It thus follows that the bulk state space can be decomposed with
respect to  the algebra 
su(2)$_L \oplus {\rm su(2)}_R \oplus \Vir_L \oplus \Vir_R$ as 
\eqn\decompb{
  \H_{\rm bulk}
= \sum_{{s,\bar s=0}\atop{s+\bar s\; {\rm even}}}^\infty 
  \left(
    V^{s/2} \otimes \bar V^{\bar s/2}\right)
  \otimes
  \left(
   \H^{\rm Vir}_{s^2/4} \otimes \bar\H^{\rm Vir}_{\bar s^2/4}
  \right)
\,.
}
The representations of the Virasoro algebra that appear in this
decomposition are all degenerate, \ie\ the corresponding Verma modules 
contain non-trivial null vectors. Indeed, for $c=1$ and $h=j^2$ with 
$j\in {1\over2}\Zop$, there is a single independent null vector at
level $(2j+1)$; if we introduce the notation 
\eqn\th{
\vartheta_s(q) = {q^{\half s^2} \over \eta(q)}
\;,\quad\ \ 
\eta(q) = q^{1\over 24} \prod_{n=1}^{\infty} (1-q^n)
}
(where $\eta(q)$ is the usual Dedekind $\eta$-function)
then the character of the Virasoro highest weight representation with
$h=j^2$ is
\eqn\char{
\chi^{\rm Vir}_h(q) 
= \vartheta_{\sqrt{2}j }(q) 
  - \vartheta_{\sqrt{2}(j+1)}(q)
\,.}
The decomposition \decompb\ implies that there are $(2j+1)^2$
Virasoro primary fields of weight $h = \bar h = j^2$ for each 
$j=0,1/2,1,...$; they transform in the $V^j \otimes \bar V^j$
representation of su(2)$_L\oplus {\rm su(2)}_R$ and are denoted 
by $\varphi^j_{m,n}(z,\bar z)$ with labels $m,n=-j,-j+1,\ldots,j$. 
These fields can be thought of as being associated to the
matrix elements of the SU(2) representation $j$, and transform in the
appropriate way under su(2)$_L\oplus {\rm su(2)}_R$. (See the Appendix
for our conventions for representations of su(2) and SU(2).) 

In order to be able to compute the partition functions (overlaps)
between two conformal boundary states we shall also need some details
about the relation between fields and the corresponding in- and 
out-states. We shall first describe our conventions in some more
generality and then concentrate on the case at hand. If we denote
the different Virasoro primary fields by $\varphi_a(z,\bar z)$ (where
in our case, $a$ stands for $(j;m,n)$), they satisfy the operator
product expansion (OPE)
\eqn\arbope{
  \varphi_a(z,\bar z) \,
  \varphi_b(w,\bar w) 
= { {g_{ab}}
  \over
   {(z-w)^{2h_a}(\bar z - \bar w)^{2\bar h_a}}
  }
 + \ldots
\,,
}
where we have only made the coupling to the identity field
explicit. Here $g_{ab}$ is some metric which we shall not assume to be   
equal to $\delta_{ab}$. (Indeed, for our case where $a =(j;m,n)$, it
is not.)  We introduce the corresponding in-states by the standard
convention 
\eqn\arbdefns{
\vec a
= \lim_{z\to 0} \varphi_a(z,\bar z) \,\vec 0
\,.
}
The out-states can then be given in terms of the action on the
out-vacuum as
\eqn\arbcev{
  \cev a
= \sum_b\;
  \lim_{z\to\infty}
  z^{2h_a} \bar z^{2\bar h_a}
  \;
 \cev 0\,g^{ab}\,\varphi_b(z,\bar z)
\,,
}
where $g^{ab}$ is the matrix inverse of $g_{ab}$. 
By definition, $\cev 0$ satisfies $\cevec 0 0 = 1$, and 
this implies then that $\cevec {a}{b} = \delta_{ab}$ as 
usual. 

If we label the fields by $(j;m,n)$, it follows from the 
su(2)$_L\oplus {\rm su(2)}_R$ symmetry that the metric is proportional
to  
\eqn\metric{
g_{(j;m,n),(j';m',n')} = \delta_{j,j'}\; 
      \delta_{m,-m'}\;\delta_{n,-n'}\ (-1)^{m-n}\,.
}
Indeed, for either su(2) the singlet is contained in the tensor
product of two representation with su(2) spins $j$ and $j'$ only if
$j=j'$; if this is the case, the singlet state is proportional (in our
conventions) to  
\eqn\singlet{
\sum_{m=-j}^{j} (-1)^m |j,m\rangle \otimes |j,-m\rangle\,.}
Applying this argument to both su(2)$_L$ and su(2)$_R$, it follows
that the metric agrees with \metric\ up to some overall (possibly
$j$-dependent) normalisation. In order to fix this, we normalise our
fields in accordance with the usual free-field construction of
\su(2)$_1$ in which the currents are given by  
\eqn\deffj{
  J^+(z) = \exp({\rm i} \sqrt 2  \phi(z)\,)
\;,\;\;
  J^3(z) = {{\rm i}
}\,\partial\phi(z) / \sqrt 2
\;,\;\;
  J^+(z) = \exp(-{\rm i} \sqrt 2  \phi(z)\,)
\,.
}
\eqn\deffjbar{
 \bar  J^+(\bar z) = \exp({\rm i} \sqrt 2  \bar \phi(\bar z)\,)
\;,\;\;
  \bar J^3(\bar z) = {{\rm i}
}\,\bar \partial\bar \phi(\bar z) / \sqrt 2
\;,\;\;
  \bar J^+(\bar z) = \exp(-{\rm i} \sqrt 2  \bar \phi(\bar z)\,)
\,.
}
In these conventions, the special Virasoro primary fields labelled 
$(j;j,j)$ and $(j;-j,-j)$ are given by 
\eqn\defff{
  \varphi^j_{j,j}(z,\bar z)
= \hat c_j \exp({\rm i}\, \sqrt 2 j (\phi(z) + \bar\phi(\bar  z)))
\,, \quad
\varphi^j_{-j,-j}(z,\bar z)
= \hat c_{-j} \exp(- {\rm i}\, \sqrt 2 j (\phi(z) + \bar\phi(\bar  z)))\,,
}
where $\hat c_j = (-1)^{2j (J^3_0 - \bar J^3_0) }$ is a cocycle needed
to ensure locality of all the fields with $j$ half-integer. 
The other Virasoro primary fields can be obtained from 
these by taking suitable contour integrals of $J^\pm$ and $\bar J^\pm$ 
around these fields. It then follows from \defff\ that 
$g_{(j;j,j),(j;-j,-j)}=1$ for all $j$, and therefore that the
normalisation in \metric\ is correct. The signs in \metric\ will pop
up later on whenever we relate in-coming and out-going boundary
states.

\subsec{Boundary conditions and structure constants}

Every boundary condition of a conformal field theory has to 
satisfy two sets of constraints: the sewing constraints
\refs{\lew} that come from considering the consistency of a
boundary conformal field theory on a genus zero world-sheet, 
and the Cardy condition  \refs{\cardy} that arises from analysing a
cylinder (\ie\ an open string one-loop diagram) with two not
necessarily equal boundary conditions. On the face of it, these
constraints are independent, although there is circumstantial evidence
suggesting that they may not be.

In this paper we shall analyse in detail one of the sewing constraints
of \refs{\lew} (that corresponds to the so-called classifying algebra
of \refs{\fs}). As we shall see, for \su$(2)_1$ this constraint is
already fairly restrictive and allows one to cut down the possible
space of conformal D-brane states quite significantly, although the
theory is not rational with respect to the Virasoro algebra. We shall
also verify that all solutions of this constraint actually satisfy the  
Cardy condition.  

Let us begin by introducing some notation. We take, without loss of
generality, the genus zero world-sheet to be the upper half-plane with
coordinates $z=x+iy$, bounded by the real line. The bulk structure
constants arise in the operator product expansions of the primary bulk
fields, 
\eqn\bulkope{
  \varphi_a(z,\bar z) \; \varphi_b(w,\bar w)
= \sum_c
  \; C_{ab}{}^c
  \;
  (z-w)^{h_c - h_a - h_b} 
  (\bar z - \bar w)^{\bar h_c - \bar h_a - \bar h_b}
  \;
  \bigl( \varphi_c(w,\bar w) + \cdots\;\bigr)
\,,
}
where, as before, the label $a$ denotes the different Virasoro primary
fields, and the ellipses correspond to higher order corrections in
$(z-w)$ or $(\bar{z}-\bar{w})$. Similarly, the boundary structure
constants arise in the operator product expansion of boundary fields 
\eqn\bope{
  \psi_i^{\alpha\beta}(x) \; \psi_j^{\beta\gamma}(y)
= \sum_k
  \; c_{ij}^{{\scriptscriptstyle (\alpha\beta\gamma)}\,k}
  \;
  (x-y)^{h_k - h_i - h_j} 
  \;
  \bigl( \psi_k^{\alpha\gamma}(y) + \cdots\; \bigr)
\,,\quad x>y\,.
}
Finally, when a primary bulk field $\varphi_a$ approaches the real
line with boundary condition $\alpha$, it can be expanded over the
fields on the boundary, defining the bulk-boundary structure constants  
$\B\alpha,a,i.$, 
\eqn\bbope{
  \varphi_a(z,\bar z)
= \sum_i 
  \; \B\alpha,a,i.
  \; (2y)^{h_i - h_a - \bar h_a}
  \; \bigl(\psi_i^{\alpha\alpha}(x) + O(y)\bigr)
}
with $z=x+iy$. Taken together, these structure constants determine the
operator algebra uniquely, and therefore all correlation functions, at
least in principle. In practice, the correlation functions are found
as sums of chiral blocks (which express the factorisation of the
correlation functions on internal channels) multiplied by the
appropriate structure constants. The chiral blocks are uniquely
determined by the representation theory of the symmetry algebra
of the boundary CFT (the Virasoro algebra, in our case). Different
possible ways to factorise on internal channels lead to different
representations for the same correlation function, and the equality of
these representations leads to constraints on the structure constants,
known as the sewing constraints.  

The constraint we are most interested in arises from considering
a two-point function of bulk primary fields on the upper half-plane,
\eqn\tpfa{
\epsfxsize 4cm
{\epsfbox[0 40 440 260]{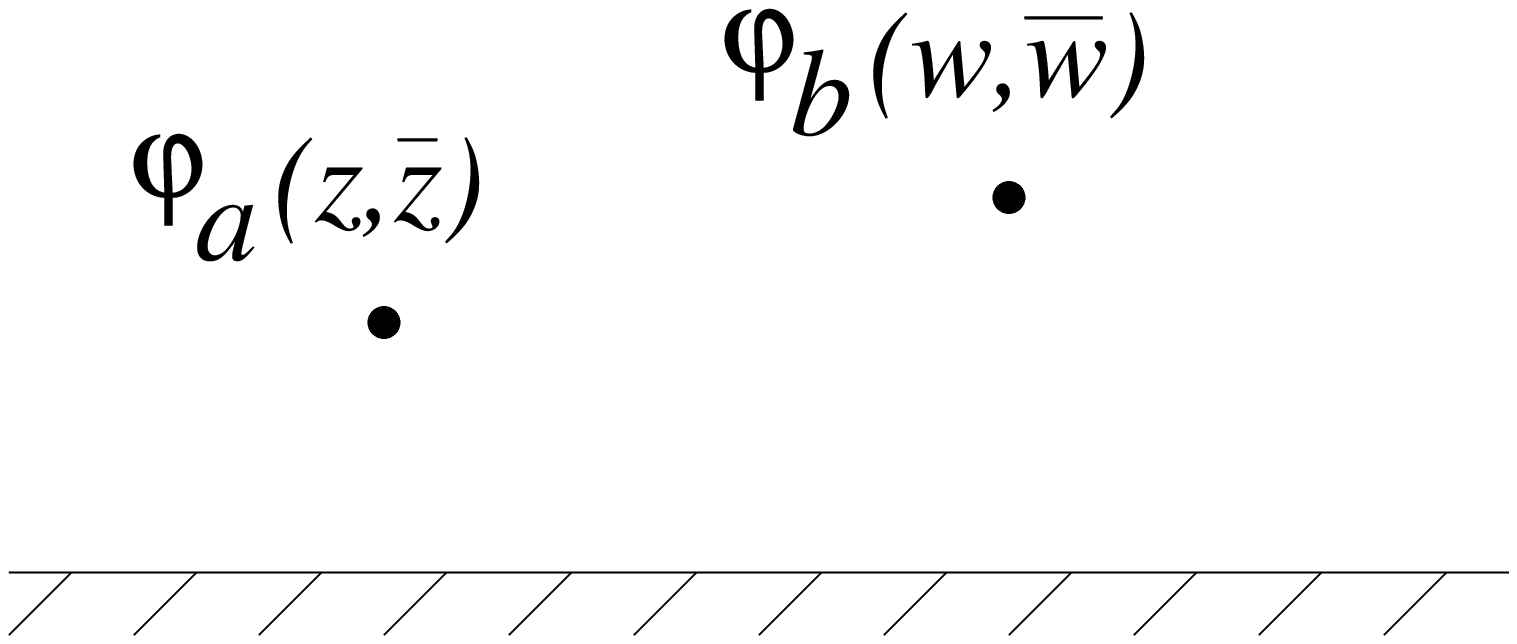}}
\;\ \ \ \ \ 
  F_{ab}(z,\bar z,w,\bar w)
= \Big\langle
  \; \varphi_a(z,\bar z) \; \varphi_b(w,\bar w) \;
  \Big\rangle
\,.
}
\vskip 3mm
The gluing conditions for the energy-momentum tensor 
imply that \tpfa\ can be described in terms of four-point chiral
blocks where we insert chiral vertex operators of weight
$h_a, \bar h_a, h_b$ and $\bar h_b$ at $z,\bar z, w$ and $\bar w$, 
respectively. 

This four-point function can be factorised in two ways, which lead to
two different representations of the correlation function, as shown
below. In the first picture, one applies the bulk-boundary OPE to  
both bulk fields first and then evaluates two-point functions of 
boundary fields. In the second picture, the bulk OPE is 
performed first, and the bulk-boundary OPE is then applied to 
the various terms in the OPE of the bulk fields. 
\eqn\cblocksa{
\epsfxsize 4cm
{\epsfbox[0 120 443 300]{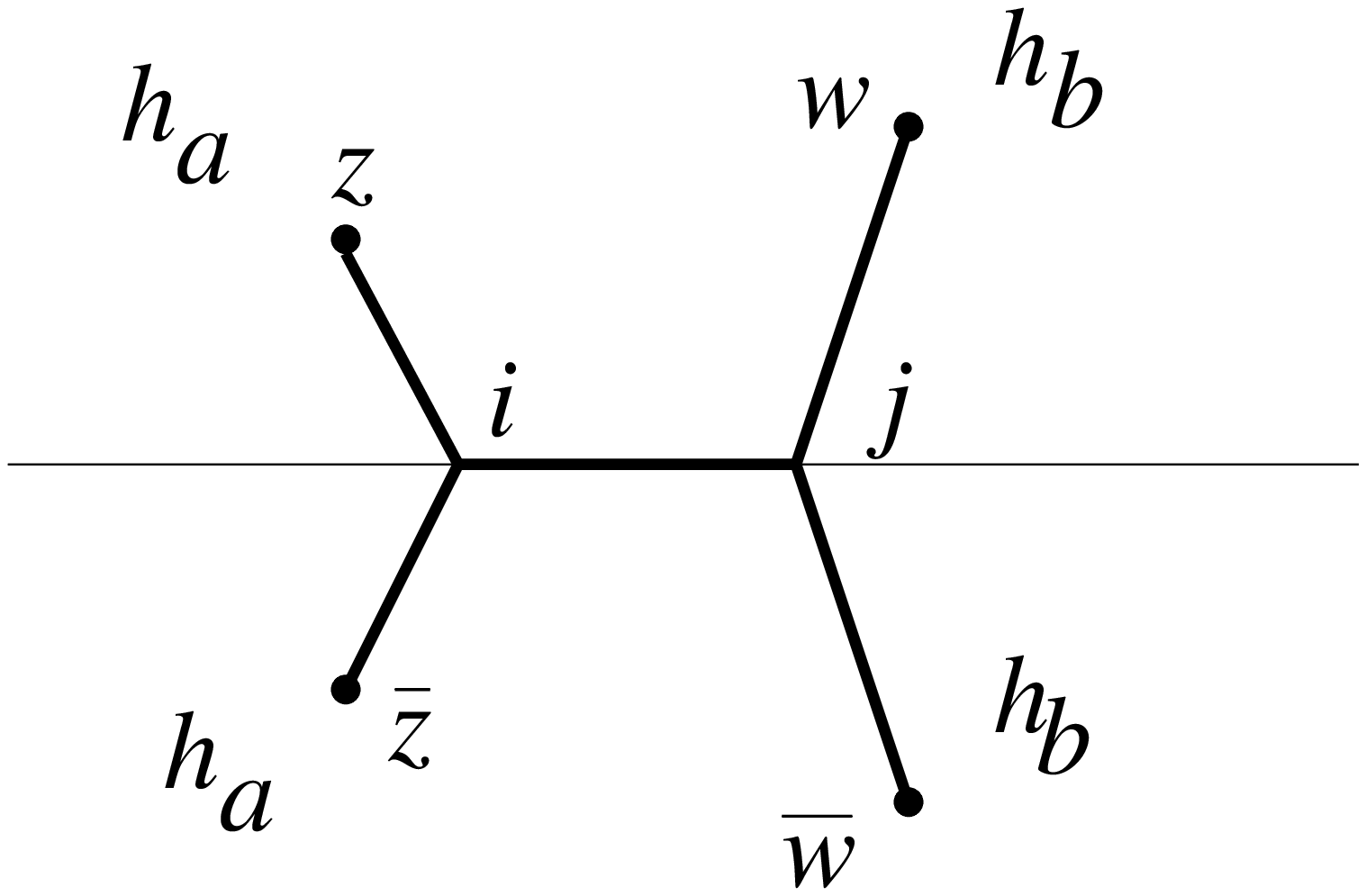}}
=
\sum_{i,j,k} \,
\left(
\B\alpha,a,i. \,
\B\alpha,b,j. \,
c_{ij}{}^k \,
\langle \psi_k \rangle\,
\right)
|z-\bar z|^{2h_b - 2h_a} \,
|z - \bar w|^{-4 h_b} \,
\cb i,a,a,b,b.(\eta)
}
\vskip 1mm
\eqn\cblocksb{
\epsfxsize 4cm
{\epsfbox[0 120 443 300]{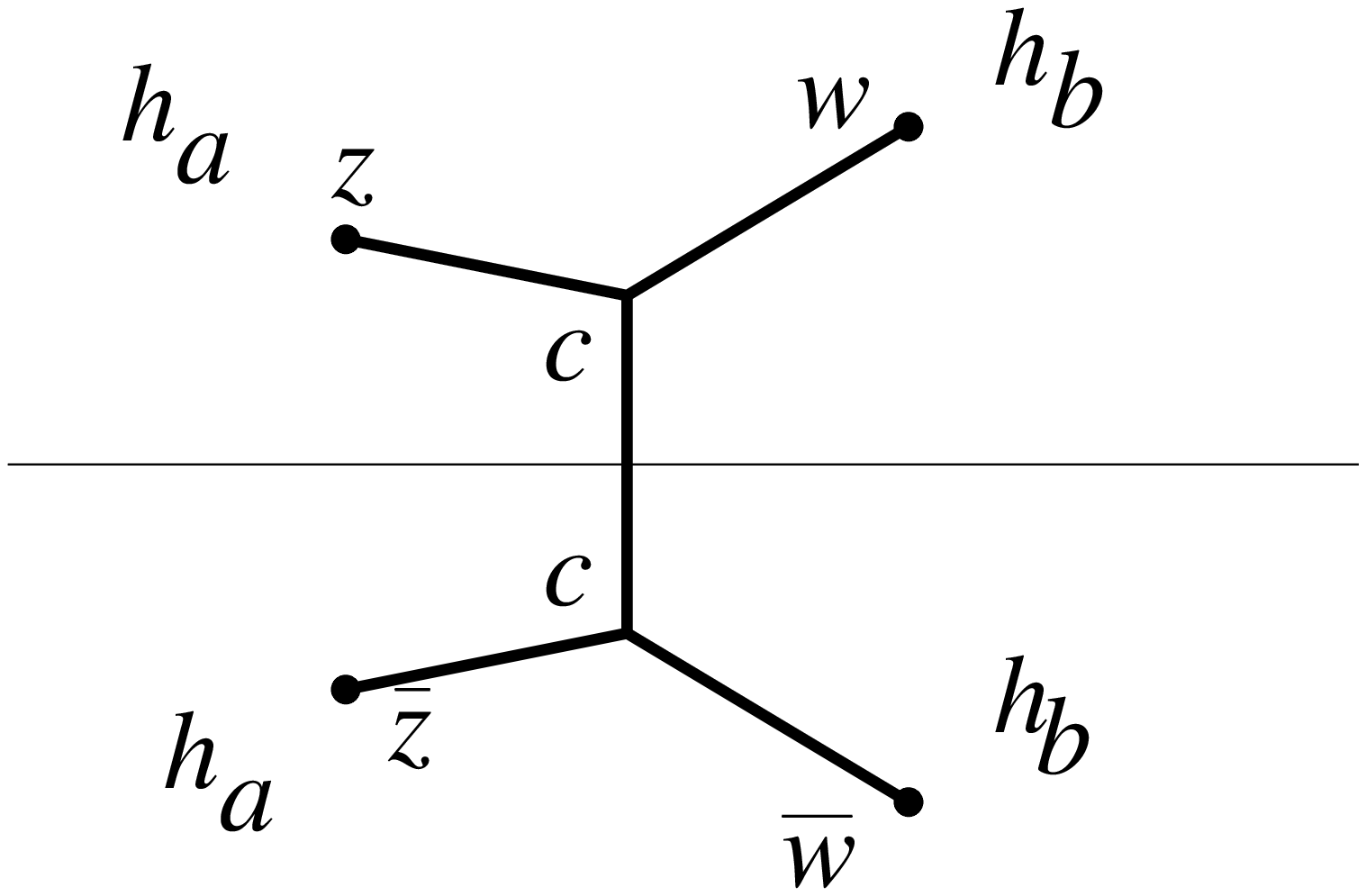}}
=
\sum_{c,k} \,
\left(
C_{ab}{}^c \,
\B\alpha,c,k. \,
\langle \psi_k \rangle\,
\right)
|z-\bar z|^{2h_b - 2h_a} \,
|z - \bar w|^{-4 h_b} \,
\cb c,a,b,b,a.(1-\eta)
}
\vskip 4mm
\noindent
In writing down these equations, we have dropped some superscripts 
labelling boundary conditions, and we have specialised to the case where  
$h_a=\bar h_a$ and $h_b=\bar h_b$. The $f^i$ and $f^c$ denote the
different chiral four-point blocks, and $\eta$ is the cross-ratio 
$\eta = |(z-w)/(z-\bar w)|^2$ which is real with $0 \leq\eta\leq 1$. 
In both equations, the $k$-summation is over all boundary fields of 
zero conformal weight, and the fields $\psi_i$ and $\psi_j$ are
necessarily of the same conformal weight; both of these statements
follow from SL$(2,\Rop)$-covariance of the boundary vacuum correlation 
functions. Normally one assumes that
there is a single bulk field of weight zero (the identity) but for 
boundary fields there are many situations in which one may want to
have a multiplicity of weight zero fields, for example to consider
superpositions of branes or to introduce Chan-Paton factors.  

The two sets of chiral blocks are related by the so-called fusing
matrices (explicit expressions for the fusing matrices of degenerate
Virasoro representations at $c=1$ relevant here are given in
\refs{\grw})  
\eqn\fusmat{
   \cb c,a,b,b,a.(1-\eta)
=  \sum_i
   \; \fm c,i,a,b,b,a.
   \; \cb i,a,a,b,b.(\eta)
\,,
}
so that substituting \fusmat\ into \cblocksb\ and comparing with
\cblocksa, one finds the sewing relation
\eqn\sewA{
\sum_{i,j} \,
\B\alpha,a,i. \,
\B\alpha,b,j. \,
c_{ij}{}^k \,
\vev{\psi_k}
=
\sum_{c} \,
C_{ab}{}^c \,
\fm c,i,a,b,b,a.
\B\alpha,c,k. 
\,
\vev{\psi_k}
\,.
}
If the algebra of weight zero boundary fields is commutative 
(and satisfies suitable further conditions, for example
semi-simplicity),  one can find a basis of projectors onto
`fundamental boundary conditions' which support a single weight zero
field (see e.g.\ \refs{\grw} for some examples). We shall now restrict
attention to such fundamental boundary conditions $\alpha$ and denote
the unique boundary field of weight 0 by $\One_\alpha$. Taking
the $k=\One$ channel in \sewA, we then arrive at the relation 
\eqn\ca{
\B\alpha,a,\One. \,
\B\alpha,b,\One. \,
=
\sum_{c} \,
C_{ab}{}^c \,
\fm c,\One,a,b,b,a.\,
\; \B\alpha,c,\One. 
\ \,.
}
Let us add some remarks on the interpretation of this relation, 
and on the mathematical structure it defines. First of all, the
condition can be recognised as the cluster condition on the
correlation functions of the boundary OPE: the projection on
$h=0$ channels that occurred in  \cblocksa\ and \cblocksb\ also arises
if the separation of the two bulk fields parallel to the boundary 
is increased, so when checking \ca\ we are investigating the `long-range
behaviour' of the correlator \tpfa, which should `cluster' as usual;
see  also \refs{\rsone}. Ignoring issues of physical interpretation,
it is tempting to regard \ca\ as the defining relations of an algebra,
with `structure constants' 
$M_{ab}{}^c \equiv C_{ab}{}^c\ \cdot \fm c,\One,a,b,b,a.$,
\eqn\cab{
  B_a \; B_b = \sum_c\  M_{ab}{}^c  \; B_c \,,
}
where, for the moment, the $B_a$ are abstract symbols. If these
relations define a commutative and associative algebra, the constraint
on the fundamental boundary conditions simply says that the 
bulk-boundary coefficients $\B\alpha,a,\One.$ have to form a
one-dimensional representation of \cab; this is the reasoning behind
calling it the `classifying algebra' \refs{\fs}. However, one should
recall that it is not yet clear whether every boundary condition that
satisfies this constraint will also satisfy the other sewing constraints, 
or Cardy's condition (and therefore whether this algebra really 
provides a one-to-one classification of the boundary conditions).   

In general, surprisingly little is known about the structure defined
by \cab. It is easy to show that the algebra is commutative since both
$C_{ab}^c$ and $\fm c,\One,a,b,b,a.$ are symmetric under the exchange
of $a$ and $b$. (Since we are only considering the Virasoro symmetry,
every primary field is self-conjugate; the symmetry of the fusing 
matrix then follows from \refs{\ms}, eq.\ (4.9).) On the other hand, 
as far as we are aware, a general proof that the algebra is 
associative is still missing. In simple cases such as rational
conformal field theories with charge-conjugate partition function and
with standard gluing conditions imposed (which, in particular,
preserve the full symmetry algebra), one can show that the classifying
algebra is nothing but the fusion ring, $M_{ab}{}^c = N_{ab}{}^c$, see 
\refs{\pss,\fs,\bppz}. Under the same assumptions, solutions to all
sewing relations were found in \refs{\fffs}, expressed in terms of the
representation category of the chiral algebra. The structure of \cab\ 
was further clarified and extended to certain non-diagonal modular 
invariants of SU(2) and Virasoro minimal models in \refs{\bppz}; for 
these cases, the $M_{ab}{}^c$ are structure constants of a Pasquier 
algebra, which in turn opens up interesting  connections to quantum 
symmetries, see \refs{\pz}.

\subsec{The boundary state formalism}

The bulk-boundary constants $\B\alpha,a,\One.$ are also an essential
ingredient in the boundary state formalism that is particularly
suited to discuss the Cardy condition. Here a boundary condition
$\alpha$ is represented by a generalised coherent state
$\iivec{\alpha}$ in the bulk theory that satisfies a number of
conditions. In order to relate boundary states with bulk-boundary 
constants, let us consider the bulk theory as being defined in the 
region outside the unit disk; the bulk correlators in the presence 
of the boundary then provide a map from the space of states, with 
field insertions outside the unit disk, to the complex numbers, and 
thus can be represented by a boundary state as 
\eqn\bdef{
  \langle\psi| \;\mapsto\; \langle\psi |\!| \alpha\rangle\!\rangle\,.
}
The condition that the boundary preserves the conformal symmetry
becomes 
\eqn\cbc{
   (L_m - \bar L_{-m})\,\iivec {\alpha}  = 0\,.}
By regarding this as an intertwiner of the left and right actions of
the Virasoro algebra, one can prove that a basis of solutions is given
by the Ishibashi states $\ivec a$ associated to bulk Virasoro primary 
fields $\varphi_a$ with $h_a = \bar h_a$ \refs{\ishi}. Thus we can
write 
\eqn\BSin{
\iivec {\alpha}
 = \sum_{(j,m,n)} A^{\alpha}_{(j;m,n)}\, \ivec{j;m,n}
\,,
}
where $A^{\alpha}_{(j;m,n)}$ are some constants (that characterise
the boundary condition). 

Alternatively, we can consider the bulk theory to be defined on the
interior of the unit disk, and describe the boundary condition in
terms of an `out'-boundary state $\iicev{\alpha}$ which satisfies 
the obvious analogue of \cbc, and for which a basis can be written in
terms of Ishibashi states $\icev a$; the general setup is
represented diagrammatically as below 
\vskip -1cm
\eqn\diskamps{
\epsfxsize 3cm
{\hskip-.25cm\epsfbox[0 80 200 260]{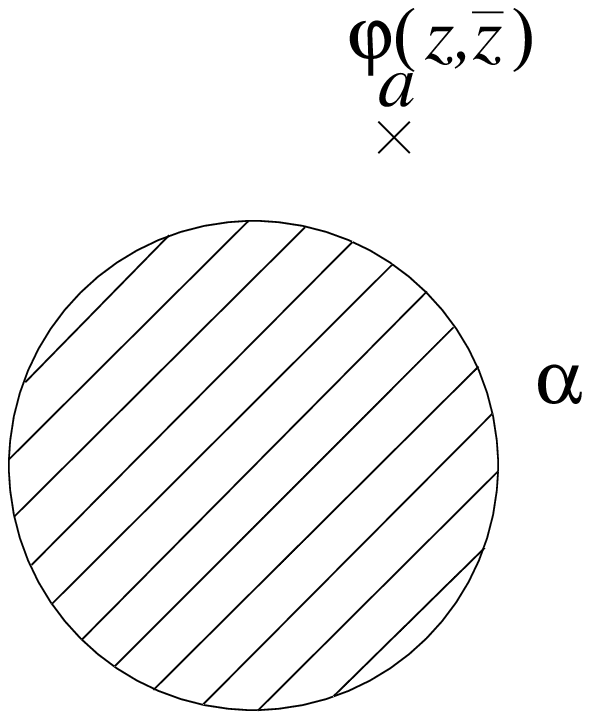}}
= \cev 0\,\varphi_a(z,\bar z)\,\iivec {\alpha}
\;,\;\;\;\;
\epsfxsize 3cm
{\epsfbox[0 100 200 280]{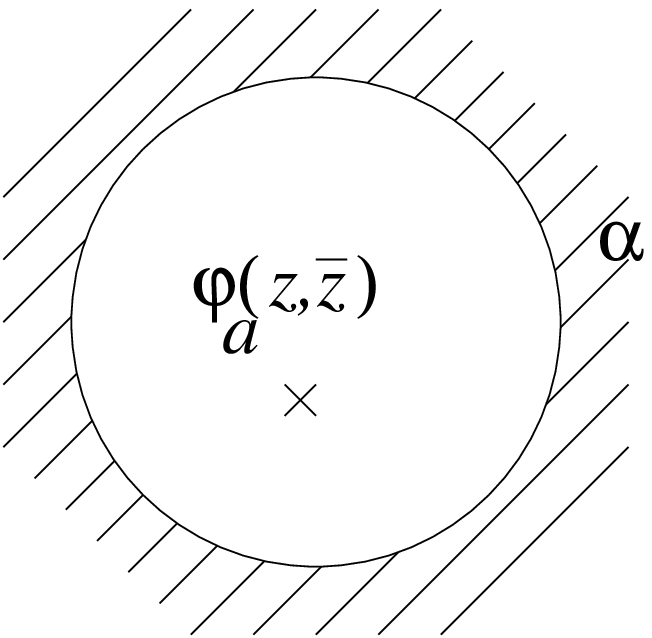}}
= \iicev{\alpha}\,\varphi_a(z,\bar z)\,\vec 0
\,.
}
\vskip 1.5cm
\noindent
If we take the boundary condition $\alpha$ to be defined by \BSin,
then the boundary state $\iicev{\alpha}$ that corresponds to the
{\it same} boundary condition $\alpha$, now viewed as `out-going', 
is given by   
\eqn\BSout{
\iicev{\alpha} =  \sum_{(j,m,n)} \sum_{(j';,m',n')} 
g_{(j;m,n),(j';m',n')} \,
A^{\alpha}_{(j;m,n)}\, \icev{j';m',n'}\,,
}
because with this convention both boundary states have the same
leading behaviour in the limit in which an arbitrary bulk field
approaches the boundary: for \BSin\ the relevant limit is 
\eqn\bbopeC{\eqalign{
{}^\alpha \!B_{(j;m,n)}^{\One}
&= \lim_{|z|\searrow 1} (|z|^2 - 1)^{2 h_j}
{{\cev 0 \varphi^j_{m,n}(z,\bar z) \iivec{ \alpha }} \over
 {\langle\, 0 \, \iivec{\alpha } }} \cr
&= {1\over \langle\, 0\, \iivec{\alpha } }
\sum_{(j,m,n)} \sum_{(j';,m',n')} g_{(j;m,n),(j';m',n')}\,
 \langle j';m',n'\iivec{\alpha} \cr
&= (-1)^{m-n}\,
   A^{\alpha}_{(j;-m,-n)} / A^{\alpha}_{(0;0,0)}
\,,
}}
where we have used the conformal symmetry of the amplitudes as well as 
\arbcev\ and the fact that (in our conventions) $g_{0,0}=1$. In the
last line we have furthermore replaced $g$ by the explicit 
expression given in \metric. This then agrees with the relevant
limit for \BSout\
\eqn\bbopeB{\eqalign{
{}^\alpha \!B_{(j;m,n)}^{\One}
&= \lim_{|z|\nearrow 1} (1 - |z|^2)^{2 h_j}
    {
 {\iicev{ \alpha } \varphi^j_{m,n}(z,\bar z) \vec 0} \over
 {\iicev{ \alpha }\, 0\, \rangle }
                 }
\cr
&=  {1\over \iicev{ \alpha }\, 0\, \rangle }
\iicev{ \alpha }\, { j;m,n}\rangle \cr
&= (-1)^{m-n} A^{\alpha}_{(j;-m,-n)} / A^{\alpha}_{(0;0,0)}
\,.}}
Incidentally, as is clear from the discussion of the previous
subsection, the relevant numerical coefficients are the bulk-boundary
structure constants that occur in \ca. 

As an aside, let us note that the same relation between in-coming and 
out-going boundary states \BSin\ and \BSout, which we have derived 
from an analysis of the bulk-boundary OPE, arises if one defines 
the out-going boundary state with the help of the CPT operator 
$\Theta$. Using that the latter acts trivially on \su(2)$_1$ Ishibashi 
states, we have
\eqn\usual{
 \iicev{\alpha} = \bigl(\, \Theta\,\iivec {\alpha}\,\bigr)^*  
}
where the star denotes the ordinary conjugation in the bulk Hilbert
space that maps $|a\rangle$ to $\langle a|$. The CPT operator was
used in the computation of open   
string amplitudes in \refs{\polcai,\rsgep}; geometrically, it arises 
because the two boundary components of the cylinder diagram 
have opposite orientations. 

Having introduced this machinery, including a careful treatment of 
the relative normalisation of Virasoro primaries, we can now discuss 
Cardy's condition
in more detail. Let us consider a cylindrical world-sheet with
boundary conditions $\alpha$ and $\beta$, where the cylinder has length
$L$ and circumference $R$. By a conformal transformation this can be
mapped to an annulus of inner radius $\exp(-2\pi L/R)$ and outer
radius 1, or to a semi-annulus on the upper half-plane of inner radius
1 and outer radius $\exp(\pi R/L)$. The corresponding partition
function can therefore be expressed in two different ways,
\eqn\partabA{
  Z_{\alpha\beta}(L,R)
= \iicev {\alpha}  q^{{1 \over 2} ( L_0 + \bar L_0 - c/12)}
  \iivec {\beta}
= {\rm Tr}_{\H_{\alpha\beta}}( \qt^{L_0 - c/24} )
\;,
}
where in the second expression $\H_{\alpha\beta}$ is the Hilbert space
for the upper half-plane with boundary condition $\alpha$ on the
positive real axis and $\beta$ on the negative real axis; furthermore, 
$ q := \exp(-4\pi L/R)$ and $\qt := \exp(-\pi R/L)$. Since the last
expression is a trace in a representation space of the
Virasoro algebra, we can decompose $Z_{\alpha\beta}(L,R)$ in terms of
irreducible Virasoro representations,
\eqn\cardyA{
    Z_{\alpha\beta}(L,R)
= \sum_{i} n_{i\alpha}{}^\beta\,\chi_i(\qt)\,.
}
Cardy's condition is then the requirement that the multiplicities
$n_{i\alpha}{}^{\beta}$ are non-negative integers. Strictly speaking,
this condition only makes sense if the spectrum of Virasoro
representations that occurs in \cardyA\ is discrete (as is necessarily
the case for rational theories). The \su$(2)_1$ model we are
considering here is only quasi-rational with respect to the Virasoro
algebra (\ie\ the model contains an infinite number of Virasoro 
representations, but each operator product of primary fields only
contains a finite number of primary fields). However, as we shall see,
the Virasoro spectrum that actually arises for the different overlaps
(corresponding to pairs of boundary states that satisfy \ca) will
always be discrete, and indeed, will always lead to non-negative
integers $n_{i\alpha}{}^\beta$. This is a strong consistency check 
on our construction.

\newsec{The general solution to the cluster condition}

Our next aim is to compute the structure constants of the 
algebra \ca\ for the case of \su$(2)_1$. The idea of the derivation is
to deduce these structure constants from the knowledge  of a
sufficiently large class of solutions.  

As we have seen above, every fundamental boundary condition (that only
preserves the conformal symmetry) has to solve \ca. While the general
solution to this problem has not been given so far (we shall present
the complete solution in the next section), a large class of boundary 
conditions that are believed to be fully consistent have been
constructed before. These boundary conditions are characterised by the
property that they preserve  the full $\su(2)$ current
symmetry. The corresponding boundary states satisfy the gluing 
conditions 
\eqn\autoB{
   \left( 
   J^a_m
 + {\rm Ad}_{(\,g \cdot\iota\,)}(\bar J^a_{-m}) 
   \right) \iivec g  = 0 
\;,\;\;\;
  m\in\Zop
\;,\;\;\;
  \iota = \pmatrix{0 & 1 \cr -1 & 0}
\;,
}
for some $g\in {\rm SU}(2)$, see \refs{\GG,\rsone}. 
(We have included the matrix $\iota$ in this definition for later
notational convenience; our conventions are such that the usual 
Cardy boundary state ${\iivec 0}_{\rm Cardy}$ associated with the 
$\su(2)_1$ vacuum representation is given by 
${\iivec 0}_{\rm Cardy}= \iivec {{\textstyle -\iota}}$. If one 
wants to avoid the $\iota$, one could work with `Neumann-like' 
Virasoro Ishibashi states $\ivec{j;m,-n}$ instead of the 
`Dirichlet-like' $\ivec{j;m,n}$ used in the following.) 
The boundary states in \autoB\ are marginal deformations of 
\su$(2)_1$ Cardy states. In terms of our Virasoro Ishibashi states, 
they are explicitly given as 
\eqn\boundsu{
   \gvec
 = {1\over 2^{1\over 4}}
  \sum_{(j;m,n)} D^j_{m,n}(g)\, 
        \ivec{j;m,n}
\;,
} 
where $D^j_{m,n}(g)$ is the matrix element of 
$g$ in the  representation $j$.

Because of \bbopeC\ the corresponding boundary
structure constants for these boundary conditions are just matrix
elements of SU(2), 
\eqn\boundsuB{
  {}^{g}B_{(j;m,n)}^{\One}
= (-1)^{m-n} \, D^j_{-m,-n}( g )
= D^j_{m,n}(\iota\cdot g\cdot \iota^{-1})
\,.}
An explicit formula for the matrix elements can be found in
\refs{\ham}  
\eqn\matrixelem{\eqalign{
D^j_{m,n}(g) & = \sum_{l=\max(0,n-m)}^{\min(j-m,j+n)}
{\left[ (j+m)!\, (j-m)!\, (j+n)!\, (j-n)! \right]^{\half} \over
(j-m-l)!\, (j+n-l)!\, l!\, (m-n+l)!} \cr
& \hskip120pt \times a^{j+n-l} d^{j-m-l} b^l c^{m-n+l}\,,}}
where we have written $g\in SU(2)$ as 
\eqn\group{
g= \pmatrix{ a & b \cr c & d} }
with $c=-b^\ast$ and $d=a^\ast$. Some useful relations involving the
matrix elements $D^j_{m,n}(g)$ have been collected in the appendix.

If we set $g=e$ in \boundsu, the boundary state describes the standard
Dirichlet brane (at $x=0$). On the other hand, standard Neumann branes
correspond to the choice $a= d =0$ in \group\ -- see the remarks after
\autoB.
\vskip8pt

If we assume that the boundary states $\gvec$ define 
{\it fundamental} boundary conditions in the sense defined
before\footnote{$^\dagger$}{While it is possible to normalise each
boundary condition $\alpha$ so that $Z_{\alpha\alpha}$ contains the
vacuum representation of the Virasoro algebra precisely once, there is
a non-trivial check that the resulting expressions for the pairwise
overlaps $Z_{\alpha\beta}$  are also given by a sum of Virasoro
characters with non-negative integer coefficients. 
As we shall see below, the above boundary states have this property
(see also \refs{\cklm,\polthor,\rsone}).} 
the different normalisation constants have to satisfy the sewing
relation \ca\ 
\eqn\factor{
  D^{j_1}_{m_1 n_1}(g) 
  \, 
  D^{j_2}_{m_2 n_2} (g)  
= \sum_{j;m,n} 
     M_{(j_1;m_1,n_1),(j_2;m_2,n_2)}{}^{(j;m,n)}
  \,
  D^{j}_{m n} (g) 
\;,}
where {\it a priori}, the coefficients $M$ on the right hand side are
products of bulk structure constants and elements of the (Virasoro) 
fusing matrix.  

The key observation is now that this family of equations (one for each
group element $g\in{\rm SU}(2)$) already determines $M$ uniquely. This
can be seen as follows: because of the Peter-Weyl Theorem (see for
example \refs{\knapp}), the matrix elements $D^j_{m,n}(g)$ define an
orthogonal basis of $L^2$ functions on the group manifold
SU(2). Since the product of two matrix elements $D^j_{m,n}(g)$
is again an $L^2$ function on SU(2), it can be
expanded uniquely in terms of these matrix elements; the expansion
coefficients are precisely the structure constants $M$, that are thus
uniquely determined by \factor.

Next we observe that \factor\ is solved if $M$ equals (see for example
\refs{\ham} eq.\ (5-116))
\eqn\Mresult{
M_{(j_1;m_1,n_1),(j_2;m_2,n_2)}{}^{(j;m,n)} = 
(j_1m_1,j_2m_2|jm)\, (jn|j_1n_1,j_2n_2)\,,}
where $(j_1m_1,j_2m_2|jm)$ and $(jn|j_1n_1,j_2n_2)$ are the
Clebsch-Gordan coefficients that describe the decomposition of the
tensor product  $j_1\otimes j_2$ in terms of the representation $j$.  
Indeed, the left hand side of \factor\ is the matrix element between
the states  labelled by $(m_1\otimes m_2)$ and $(n_1\otimes n_2)$ in
the tensor product of the representations $j_1$ and $j_2$; the
Clebsch-Gordan coefficients describe the decomposition of this tensor
product into irreducible representations, and therefore these matrix
elements must agree with the right hand side of \factor. Since the 
structure constants $M$ are uniquely determined by \factor, we may 
conclude that $M$ must be given by \Mresult.   

The derivation of \Mresult\ is somewhat indirect, and indeed assumes
that the familiar set of boundary states $\gvec$ for \su$(2)_1$  
actually define consistent boundary conditions. As a consistency check
on our analysis we have in a few cases verified by explicit
computation that the constants $M$ so derived agree with the formula
in terms of the bulk structure constants and the fusing matrices.

\subsec{The general solution}

As we have seen before, the normalisation constants of every
fundamental D-brane define a one-dimensional representation of
the algebra \ca. Now that we have identified the structure constants
of this algebra, we can classify all its one-dimensional
representations, and thus obtain an `upper bound' on the set of all  
fundamental branes of the theory, irrespective of whether
these boundary conditions preserve \su(2)$_1$ or just the Virasoro
algebra.  

Let us denote by $B^j_{m,n}={}^\alpha \!B_{(j;m,n)}^{\One}$ 
the relevant bulk-boundary structure constants for such a 
general boundary state labelled $\alpha$. We are looking for the 
most general solution to the (non-linear) equation
\eqn\factorone{
  B^{j_1}_{m_1,n_1} B^{j_2}_{m_2,n_2} 
= \sum_{j;m,n} 
  M_{(j_1;m_1,n_1),(j_2;m_2,n_2)}{}^{(j;m,n)}
  B^{j}_{m,n} 
\,.}
If the solution is non-trivial (\ie\ if at least one $B^{j}_{m,n}\ne 0$), 
then we have to have $B^0_{0,0}=1$, since the Clebsch-Gordan
coefficients satisfy $(j'n'|jn,00)=\delta_{jj'} \delta_{nn'}$.
Next we observe that once we have chosen  
\eqn\solone{
  B^{\half}_{m,n} 
= \pmatrix{a & b \cr c & d}_{m,n}
\,,}
all other coefficients $B^j_{m,n}$ are uniquely determined (if they
can be consistently found); this is simply a consequence of the fact
that every representation of SU(2) is contained in multiple tensor
products of the $j=\half$ representation. It thus remains to analyse
what consistency condition on \solone\ guarantees that a solution for
all $B^j_{m,n}$ can be obtained.

One such consistency condition can be easily found: using the explicit
form of the Clebsch-Gordan coefficients, see e.g.\ eq.\ (9-122) of 
\refs{\ham}, we have that 
\eqn\condone{\eqalign{
    B^\half_{-\half,\half} B^\half_{\half,-\half} 
& = \half B^1_{0,0} - \half B^0_{0,0}\;, 
\cr
\noalign{\vskip6pt}
    B^\half_{-\half,-\half} B^\half_{\half,\half} 
&= \half B^1_{0,0} + \half B^0_{0,0}\;.
}}
This implies, in particular, that the matrix in \solone\ must have 
determinant one: 
\eqn\condtwo{
  B^\half_{-\half,-\half} B^\half_{\half,\half} 
- B^\half_{-\half,\half} B^\half_{\half,-\half} 
= ad - bc
= B^0_{0,0} = 1\,. }
We shall now argue that this is, in fact, the only consistency
condition. It is clear from the structure of \factorone\ that the
consistency conditions are polynomial relations in $a,b,c,d$, and
therefore, that the relations do {\it not} involve the complex
conjugate of any of these coefficients. Furthermore, if \solone\ is a
matrix in SU(2), then the consistency conditions are manifestly
satisfied. However, the only polynomial relation that is actually
satisfied by every SU(2) matrix is the condition that the
determinant is equal to one. Thus \condtwo\ must be the only
consistency condition. We therefore conclude that the (fundamental)
conformal boundary conditions of \su$(2)$ at level $k=1$ are (at most) 
parametrised by group elements in SL$(2,\Cop)$.

\newsec{Cardy condition and the boundary spectrum revisited}

In this section we want to check whether the family of boundary states
that are associated to group elements in SL$(2,\Cop)$ actually
satisfy Cardy's condition. To this end we need to calculate the
various overlaps between the in- and out-boundary states we have 
defined before.  

The annulus partition function $\A\equiv Z_{g_1,g_2}$ for an annulus
with boundary condition $g_1$ on the outer circumference and
$g_2$ on the inner one is given in terms of boundary states as 
\eqn\apf{\eqalign{
       \A 
&=     \iicev { g_1 } 
   \,  q^{\half(L_0+\bar{L}_0-{c\over12})}
   \,  \iivec { g_2 }
\cr
&= 
{1\over \sqrt{2}} 
\sum_{j\in\half\Zop_+} \sum_{m,n} 
(-1)^{m-n}D^j_{-m,-n}(g_1) 
D^j_{m,n}(g_2)\, \chi_{j^2}(q) \,. \cr}}
Because of (A.6) the sum over $m$ and $n$ simplifies to
\eqn\calcone{
\sum_{m,n} D^j_{n,m} (g_1^{-1}) D^j_{m,n}(g_2) = 
\sum_{n} D^j_{n,n}(g_1^{-1} g_2) \,,}
where we have used the property that the matrix elements $D^j_{m,n}$
also define a representation for SL$(2,\Cop)$. (Again, this follows 
from the fact that the representation property is a polynomial relation 
in $a,b,c,d$, and that the only polynomial relation that is satisfied by
SU(2) is $ad-bc=1$.) Next we observe that \calcone\ is simply the
trace in the $j^{{th}}$ representation of $g=g_1^{-1}g_2$. By
conjugation by an element of SL$(2,\Cop)$ (that does not modify the
value of the trace), we can bring $g$ into `Jordan normal form'
$\hat{g}$ (\ie\ we can choose $\hat{c}=0$); the matrix elements
$D^j_{n,n}(\hat{g})$ then simplify to 
\eqn\calctwo{
D^j_{n,n}(\hat{g}) = \hat{a}^{j+n} \hat{d}^{j-n} = \hat{a}^{2n} \,,}
where we have used that $\hat{a}\hat{d}=1$ since $\hat{g}$ has
determinant one. The sum over $n$ can now be easily performed, and we
find 
\eqn\calcthree{ \sum_{n} D^j_{n,n}(g_1^{-1} g_2) = 
{\sinh((2j+1)\alpha) \over \sinh(\alpha)}\,,}
where $\hat{a}=e^{\alpha}$.

Next, we rewrite the Virasoro character $\chi_{j^2}$ in terms of the
functions $\vartheta$ as in \char, and thus obtain
\eqn\calcfour{\eqalign{
\sqrt{2} \A & = \sum_{j\in\half\Zop_+} 
{\sinh((2j+1)\alpha) \over \sinh(\alpha)} 
\left(\vartheta_{\sqrt{2}j}(q) - \vartheta_{\sqrt{2}(j+1)}(q) \right) \cr
& = \vartheta_0 (q) + \sum_{j=1}^{\infty} \vartheta_{\sqrt{2}j}(q)
\left({\sinh((2j+1)\alpha) \over \sinh(\alpha)} - 
{\sinh((2j-1)\alpha) \over \sinh(\alpha)} \right) \cr
& \quad + 2 \cosh(\alpha)\, \vartheta_{1\over\sqrt{2}} (q) 
+ \sum_{j={3\over 2}}^{\infty} \vartheta_{\sqrt{2}j}(q)
\left({\sinh((2j+1)\alpha) \over \sinh(\alpha)} - 
{\sinh((2j-1)\alpha) \over \sinh(\alpha)} \right) \cr
& = \sum_{j\in\half\Zop} \cosh(2j\alpha)\, \vartheta_{\sqrt{2}j}(q) \,,}}
where the sum in the second (third) line is over integers (half-odd
integers) only, and where we have used the identity 
\eqn\trigo{
{\sinh((2j+1)\alpha) \over \sinh(\alpha)} - 
{\sinh((2j-1)\alpha) \over \sinh(\alpha)} = 2 \cosh(2j\alpha) }
as well as the fact that $\vartheta_s(q)$ only depends on $|s|$. Under
a modular transformation, the $\vartheta$-functions transform as  
\eqn\modular{
\vartheta_s(q) = \int_{-\infty}^{\infty} dt\; e^{2\pi i t s} \,
       \vartheta_t(\tilde{q})\,,}
where $\tilde{q}$ is the `open string' parameter.  (More precisely, if
we write $q=e^{2\pi i\tau}$ with 
${\rm Im}\,\tau>0$,
then $\tilde{q}=e^{-{2\pi i \over \tau}}$.) Putting these results 
together we then find that in the open string description the overlap 
becomes 
\eqn\calcfive{\eqalign{
\A & = {1\over \sqrt{2}} 
\int_{-\infty}^{\infty} dt\, \vartheta_t(\tilde{q})
\sum_{j\in\half\Zop} e^{2\pi i \sqrt{2} j t} \cosh(2j\alpha)  \cr
& = {1\over \sqrt{2}} 
\int_{-\infty}^{\infty} dt\, \vartheta_t(\tilde{q})
\sum_{l\in\Zop} e^{i l (\sqrt{2}\pi t - i \alpha)} \cr
& = {2 \pi\over \sqrt{2}} 
\int_{-\infty}^{\infty} dt\, \vartheta_t(\tilde{q})
\sum_{n\in\Zop} \delta(\sqrt{2}\pi t - i \alpha + 2 \pi n) \cr
& = \sum_{n\in\Zop} 
\vartheta_{{i\alpha\over \sqrt{2}\pi} + \sqrt{2}n}(\tilde{q})\,.}}
For all values of $\alpha$ this defines a positive integer linear 
combination of Virasoro characters. 

Since $\alpha$ only depends on ${\rm Tr}(g_1^{-1} g_2)$, the open
string spectrum between any of the above D-branes (labelled by  
$g_1\in {\rm SL}(2,\Cop)$) and itself is the same, irrespective of
$g_1$. (For the case when $g_1\in {\rm SU}(2)$, this was observed
before in \refs{\cklm,\polthor,\rsone}.) In fact, in this case
$\alpha=0$, and we therefore obtain only Virasoro representations for
which $h$ is  non-negative. Furthermore, the vacuum representation
(\ie\ the character with $h=0$) occurs precisely once, as was claimed
previously for the case with $g_1\in {\rm SU}(2)$. Indeed, the fact
that this is true for all $g_1\in {\rm SL}(2,\Cop)$ is necessary for
the consistency of our  approach: the constraint that a boundary state
has to provide a representation of the algebra \ca\ only arises for
`fundamental' D-branes, \ie\ for those boundary states for which the
vacuum representation appears precisely once in the open string
spectrum. 

For the case where $g=g_1^{-1} g_2\in {\rm SU}(2)$, $\alpha$ is purely
imaginary, and thus all Virasoro highest weights that occur in the
open string  between two such D-branes are positive. However, in
general also negative or even complex values for the conformal weights
occur in the open string partition function. In particular, if $g_1$
is an arbitrary  element in SL$(2,\Cop) {\hsmallsetminus} {\rm SU}(2)$,
we can always find $g_2\in {\rm SU}(2)$ so that the overlap between
$g_1$ and $g_2$ leads to imaginary conformal weights in the open
string.

\newsec{Discussion}

Let us recapitulate what we have shown in this paper. Under the
assumption that the well-known boundary states that are marginal 
deformations of \su(2)$_1$ Cardy boundary states and are parametrised by 
$g\in {\rm SU}(2)$ define consistent boundary conditions, we have 
derived the structure constants of the `classifying algebra' \ca. 
We have then shown that the most general one-dimensional
representation of \ca\ can be described in terms of group elements in
SL$(2,\Cop)$. Since every fundamental D-brane has to satisfy this
condition, this implies that the complete set of fundamental conformal
boundary conditions for \su$(2)_1$ is contained in this family of
boundary states. Naively, one might have expected that the space 
of conformal boundary conditions (which preserve the Virasoro algebra 
only and render the boundary theory non-rational) is much larger.

One can ask whether the branes parametrised by 
SL$(2,\Cop){\hsmallsetminus} {\rm SU}(2)$ are indeed `new' in the sense
that they cannot be written as superpositions of the boundary states
associated to elements in SU(2). Since we are dealing with a continuum
of boundary states here, this is not a purely algebraic question, but
some analytic considerations come into play. It is natural to believe
that the relevant superpositions are of the form   
\eqn\fourier{\eqalign{
|\!| B\,\rangle\!\rangle = 
  \int_{\rm SU(2)} d\mu(g)\; F_{B}(g) \;\gvec\,,}
}
where $F_B$ are tempered distributions on the manifold SU(2) with Haar
measure $d\mu$. (For example, $F_B$ is a delta-function if 
$|\!| B\,\rangle\!\rangle$ is taken from our SU(2) family of boundary
states.) The boundary states associated to 
SL$(2,\Cop) {\hsmallsetminus} {\rm SU}(2)$ cannot be written in this
form, and are therefore likely to be genuinely new.
\vskip4pt

As we have stressed before, it is {\it a priori} not clear whether all
of these boundary states actually define consistent boundary
conditions. In particular, while all of these more general boundary
states satisfy Cardy's condition, we have not checked whether they
satisfy all other sewing constraints. 

In general, we have little to say about this problem. However, there
exist various subclasses of boundary states which, very plausibly
define consistent boundary conditions. In particular, we can consider
the cosets of SU(2) in SL$(2,\Cop)$. Since the overlap between
two boundary states depends only on ${\rm Tr}(g_1^{-1} g_2)$, it
follows that the relative overlaps of this subset of D-branes are
precisely the same as those of the original SU(2) family. Thus the
boundary field content is the same and therefore these branes
presumably define consistent boundary conditions. In fact, if we write
$g=h\cdot u$ where $h$ is a fixed element $h  \in {\rm SL}(2,\Cop)$ that
characterises the coset, and $u\in {\rm SU}(2)$ is arbitrary, we can
relate the corresponding boundary states as
\eqn\newbsB{
 \iivec{ h\cdot u }
= \exp(\theta_a \bar J{}^a_0)
  \,
  \iivec{ u }
\; ,
}
using a representation $h= \exp(\theta_a t^a)$ in terms of Lie algebra 
generators. 
It then follows that the boundary conditions associated to different
cosets define completely equivalent field theories since they can be
related by conjugation of the bulk fields by $h$. 

While (at least) these cosets of boundary states seem to define
consistent boundary conditions from the point of view of conformal
field theory, the corresponding D-branes are presumably not of
interest in string theory (except, of course, for the original
D-branes that are associated to group elements in SU(2)). Indeed,
the simplest example of a group element in
SL$(2,\Cop){\hsmallsetminus}{\rm SU}(2)$ is   
\eqn\newg{
 g = \pmatrix{ \exp a & 0 \cr 0 & \exp ( - a ) }
\,,
}
where $a\ne 0$ is real. In the free-field construction, the
corresponding D-brane can be thought of as a Dirichlet brane whose
position takes a purely {\it imaginary} value. In general, one can
similarly show that group elements in 
SL$(2,\Cop){\hsmallsetminus}{\rm SU}(2)$ lead to D-branes that have 
imaginary couplings to some of the bulk fields. If we discard these
`unphysical' boundary conditions then the results of our paper 
imply that all `physical' D-branes of the $\su(2)_1$ theory
preserve actually the full SU(2) symmetry! 

The problem of classifying all conformal boundary conditions of a
single free boson at the self-dual radius was discussed by Friedan in
unpublished work \refs{\friedan}. By the Frenkel-Kac-Segal
construction this theory is equivalent to the SU$(2)$ WZW model at
$k=1$ that we have been considering in this paper. Friedan's paper
contains a bare outline of some of the steps in the calculation, and
then claims that the space of conformal boundary conditions is
SU$(2)$. Since he only provides very few details about the properties
he requires of a  conformal boundary condition, it is not possible to
say quite why he finds this space, and not SL$(2,\Cop)$ as we do. One
possible interpretation is that he has implicitly assumed that 
the in- and out- boundary states are related as in \usual, without
the CPT operator $\Theta$. Alternatively, he may have discarded these
additional boundary states because of their unphysical properties from
a string theory point of view. However, since his text does not 
contain any statement about these matters, it is impossible to make 
a thorough comparison with our results. 
\vskip4pt

Our analysis is quite specific to the SU(2) WZW at level $k=1$ and
does not directly generalise to $k>1$. Indeed, as was shown recently
in \refs{\mms}, the \su$(2)_k$ theory for $k>1$ possesses (physical)
D-branes that do not preserve the affine symmetry.  Extensions to
other examples of truly symmetry breaking boundary conditions
(rendering the boundary theory non-rational) require a relatively
detailed understanding of how representations of the chiral algebra of
the bulk theory decompose into those of the smaller symmetry
algebra. On the other hand, the results of this paper should
generalise fairly directly to the case of \su$(n)$ at $k=1$ where, for
$n>2$, the Virasoro algebra is replaced by $W_n$, the Casimir algebra
of $\su(n)_1$. (The W-algebra $W_n$ is the commutant of su$(n)$ in
$\su(n)$.) We then expect that D-branes that respect the $W_n$
symmetry are parametrised by elements in SL$(n,\Cop)$, rather than
just by elements in SU(n). This would provide another class of
examples where there are far fewer symmetry breaking boundary
conditions than one may have naively thought.
\vskip1cm

\centerline{{\bf Acknowledgements}}\pano

\noindent We thank Brian Davies, Romek Janik, Juan Maldacena,
Daniel Roggenkamp, Katrin Wendland and Jean-Bernard Zuber for useful
discussions, and  Jean-Bernard Zuber for  providing us with a copy of
Friedan's unpublished manuscript. M.R.G.\ is grateful to the Royal
Society for a University Research Fellowship, and the research of
A.R.\ is supported  in part by the Nuffield Foundation, Grant Number
NUF-NAL/00421/G.  

\vskip.8cm

\appendix{A}{Some properties of matrix elements}

In this appendix we shall collect some useful identities involving 
the matrix elements \matrixelem\ of SU(2) and fix our conventions for
the action of su(2) generators. First of all, since the coefficients
of the sum in \matrixelem\ are real, we have
\eqn\appone{
\left\{D^j_{m,n}\left[\pmatrix{a & b \cr c & d} \right]\right\}^\ast = 
D^j_{m,n}\left[\pmatrix{a^\ast & b^\ast \cr c^\ast &  d^\ast}
\right]\,.}
Next we observe that 
\eqn\apptwo{
D^j_{-n,-m}\left[\pmatrix{a & b \cr c & d} \right] = 
D^j_{m,n}\left[\pmatrix{d & b \cr c & a} \right]\,,}
as follows directly from the definition of \matrixelem. Similarly we
find that 
\eqn\appthree{
(-1)^{m-n} D^j_{m,n}\left[\pmatrix{a & b \cr c & d} \right] = 
D^j_{m,n}\left[\pmatrix{a & -b \cr -c & d} \right] \,.}
Furthermore, by writing out the left hand side and changing the
summation variable from $l$ to $\hat{l}=n-m+l$ we derive that 
\eqn\appfour{
D^j_{n,m}\left[\pmatrix{a & b \cr c & d} \right] = 
D^j_{m,n}\left[\pmatrix{a & c \cr b & d} \right]\,.}
Taking \apptwo, \appthree\ and \appfour\ together we then obtain
\eqn\appfive{
(-1)^{m-n} D^j_{-m,-n}\left[\pmatrix{a & b \cr c & d} \right] = 
D^j_{m,n}\left[\pmatrix{d & -c \cr -b &  a} \right]\,.}
This implies that, for all $g \in {\rm SL}(2,\Cop)$, 
we have 
\eqn\appeight{ 
(-1)^{m-n}D^j_{-m,-n}(g) = D^{j}_{n,m}(g^{-1}) \;.}
For the case where $g\in SU(2)$, we can use $c=-b^\ast$ and $d=a^\ast$ 
to show that 
\eqn\appsix{
\left[D^j_{m,n}(g)\right]^\ast = (-1)^{m-n} D^j_{-m,-n}(g)\,.}
Finally, the action of su(2) in the representation $V^j$ with basis 
$\vec{j,m}$ is determined by \matrixelem\ to be 
\eqn\sutwodef{
\eqalign{
  J^3\,\vec {j,m} &= m\,\vec {j,m} \;,\;\;
\cr
  J^+\,\vec {j,m} &= \sqrt{(j-m)(j+m+1)}\,\vec {j,m+1}\;,\;\;
\cr
  J^-\,\vec {j,m} &= \sqrt{(j+m)(j-m+1)}\,\vec {j,m-1}\;.
}}
\vskip1.6cm

\footatend%\vfill\supereject
\immediate\closeout\rfile\writestoppt
\baselineskip=14pt\centerline{{\bf References}}\bigskip{\frenchspacing%
\parindent=20pt\escapechar=` \input refs.tmp\vfill\eject}\nonfrenchspacing

\bye